\documentclass[conference]{IEEEtran}
\IEEEoverridecommandlockouts
\usepackage{cite}
\usepackage{amsmath,amssymb,amsfonts}
\usepackage{algorithmic}
\usepackage{graphicx}
\usepackage{textcomp}
\usepackage{xcolor}
\usepackage[colorlinks, urlcolor=black, linkcolor=blue, citecolor=blue]{hyperref}
\usepackage{multirow}
\usepackage{array}
\usepackage{makecell}
\usepackage{placeins}
\def\BibTeX{{\rm B\kern-.05em{\sc i\kern-.025em b}\kern-.08em
    T\kern-.1667em\lower.7ex\hbox{E}\kern-.125emX}}
\begin{document}

\title{Learning Generalizable Features for\\Tibial Plateau Fracture Segmentation Using\\Masked Autoencoder and Limited Annotations}

\author{\IEEEauthorblockN{Peiyan Yue$^{1,\dagger}$, Die Cai$^{2,\dagger}$, Chu Guo$^{4}$, Mengxing Liu$^{3,4}$, Jun Xia$^{2,*}$, Yi Wang$^{1,*}$
\thanks{$\dagger$ Peiyan Yue and Die Cai contribute equally to this work.}
\thanks{$*$ Corresponding authors: Jun Xia and Yi Wang.}
\thanks{This work was supported in part by
the Guangdong-Hong Kong Joint Funding for Technology and Innovation under Grant 2023A0505010021,
the National Natural Science Foundation of China under Grants 62471306, 62071305 and 82171913,
the Shenzhen Medical Research Fund under Grant D2402010,
the Shenzhen Science and Technology Program (JCYJ 20220818101816036),
the Shenzhen Second People's Hospital Key Clinical Research Project (20243357010),
and the Guangdong Basic and Applied Basic Research Foundation under Grant 2022A1515011241.}
}
\IEEEauthorblockA{
$^{1}$Smart Medical Imaging, Learning and Engineering (SMILE) Lab,
Medical UltraSound Image Computing (MUSIC) Lab,\\
School of Biomedical Engineering, Shenzhen University Medical School, Shenzhen University, Shenzhen, 518060, China\\
$^{2}$Department of Radiology, The First Affiliated Hospital of Shenzhen University, Shenzhen University,\\
Shenzhen Second People's Hospital, Shenzhen, 518035, China\\
$^{3}$Shenzhen Mindray Bio-Medical Electronics Co., Ltd, Shenzhen, 518132, China\\
$^{4}$Wuhan Mindray Scientific Co., Ltd, Wuhan, 430010, China
}
}

\maketitle

\begin{abstract}
Accurate automated segmentation of tibial plateau fractures (TPF) from computed tomography (CT) requires large amounts of annotated data to train deep learning models, but obtaining such annotations presents unique challenges.
The process demands expert knowledge to identify diverse fracture patterns, assess severity, and account for individual anatomical variations, making the annotation process highly time-consuming and expensive.
Although semi-supervised learning methods can utilize unlabeled data, existing approaches often struggle with the complexity and variability of fracture morphologies, as well as limited generalizability across datasets.
To tackle these issues, we propose an effective training strategy based on masked autoencoder (MAE) for the accurate TPF segmentation in CT.
Our method leverages MAE pretraining to capture global skeletal structures and fine-grained fracture details from unlabeled data, followed by fine-tuning with a small set of labeled data.
This strategy reduces the dependence on extensive annotations while enhancing the model's ability to learn generalizable and transferable features.
The proposed method is evaluated on an in-house dataset containing 180 CT scans with TPF.
Experimental results demonstrate that our method consistently outperforms semi-supervised methods, achieving an average Dice similarity coefficient (DSC) of 95.81\%, average symmetric surface distance (ASSD) of 1.91mm, and Hausdorff distance (95HD) of 9.42mm with only 20 annotated cases.
Moreover, our method exhibits strong transferability when applying to another public pelvic CT dataset with hip fractures, highlighting its potential for broader applications in fracture segmentation tasks.
Our code is publicly available at
\textit{\textcolor{blue}{https://github.com/yuepeiyan/TPFSeg-MAE}}.
\end{abstract}


\section{Introduction}
Tibial plateau fractures (TPF) present a significant challenge in medical imaging due to their complexity and variability~\cite{tscherne1993tibial}.
The detection and analysis of these fractures are further complicated by their specific characteristics and diverse presentations, making accurate diagnosis labor-intensive for clinicians.
Automated segmentation methods based on deep learning hold promise for addressing these issues, yet their effectiveness relies heavily on large and annotated datasets that are challenging to obtain.
The acquisition of such data is limited by patient privacy concerns, medical ethics, and their requirement for expert medical annotations. 
Annotating TPF from computed tomography (CT) is particularly demanding, as it requires precise identification of fracture patterns, evaluation of severity, and consideration of individual anatomical variations, all within images often affected by noise, artifacts, and inconsistent imaging parameters.
These challenges make the creation of high-quality annotated datasets both costly and time-consuming, underscoring the need for segmentation methods that perform effectively with limited annotated data.

\begin{figure}[t]
	\begin{center}
		\includegraphics[width=\linewidth]{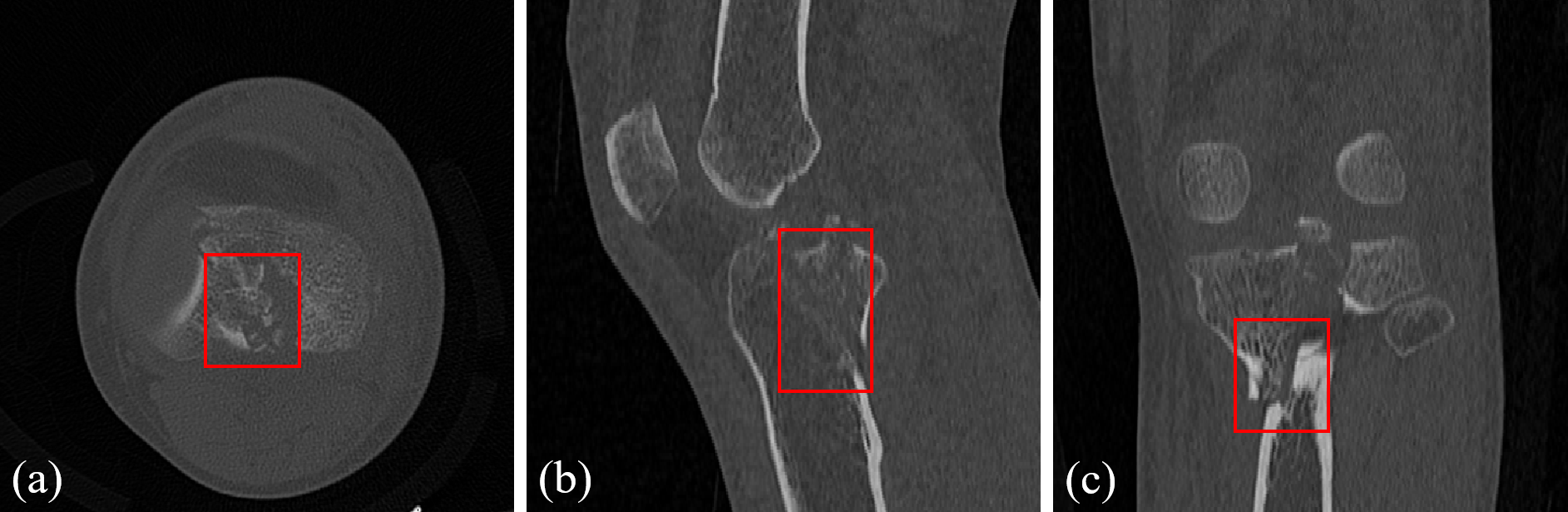}
	\end{center}
	\caption{Illustration of tibial plateau fractures (TPF) in computed tomography (CT) slices from three different views: (a) axial view, (b) sagittal view, and (c) coronal view.
	The fractures exhibit considerable variability in location and morphology, with some fragments being particularly difficult to distinguish, presenting huge challenges for the accurate segmentation.}
	\label{fig:challenge}
\end{figure}

Semi-supervised learning, which utilizes a small set of labeled images in combination with a large pool of unlabeled images, has been successfully applied to various bone segmentation tasks~\cite{zhao_2019_SemiSupervised,burton_2020_Semisupervised,liu_2020_Semisupervised,li_2023_Patchshufflebased,lu_2024_Better,li_2025_Semisupervised}.
However, these semi-supervised learning methods still face limitations in TPF segmentation tasks.
Firstly, existing semi-supervised learning methods for bone segmentation often rely on consistency regularization or self-training approaches~\cite{tarvainen_2018_Mean, berthelot_2019_MixMatch, zeng2021reciprocal, bao_2024_Robust, yang2024non, bi_2024_Sampleweighted, chen_2024_Virtual, zhong2025bounding}.
These methods typically use predictions generated by a teacher model on unlabeled data to supervise the training of a student model. 
However, as shown in Fig.~\ref{fig:challenge}, the diverse types and complex fragment morphologies of TPF pose huge challenges.
Models trained on a small set of finely annotated data struggle to generate high-quality predictions for these highly variable fracture regions, ultimately affecting the overall training performance. 
Secondly, models trained using semi-supervised learning methods often exhibit satisfactory performance only on specific datasets, making it difficult to generalize or transfer their effectiveness to other datasets.

In this study, we propose a pretrain-finetune strategy based on masked autoencoder (MAE) to improve performance of the TPF segmentation task with limited annotations.
First, the MAE is used to perform an image reconstruction task on unlabeled data, enabling the model to learn the anatomical prior of bones and subtle fracture characteristics.
Then, the model is fine-tuned with a small set of labeled data to guide the segmentation process.
This training strategy reduces the reliance on annotated data and allows the network to capture more general and transferable fracture features, enhancing its effectiveness in handling diverse fracture cases.
We evaluate the efficacy of our method on an in-house dataset, which contains 180 CT scans with TPF.
Additionally, we further employ a public dataset containing 103 pelvic CT scans with hip fractures to assess the transferability of our method.
Experimental results demonstrate that our method achieves favorable segmentation accuracy on TPF tasks with limited annotations, and shows promising transferability across different datasets.

\section{Related Work}
In recent years, several studies have focused on the automated segmentation of bone structures.
Besler~\textit{et al}.~\cite{besler_2021_Bone} proposed an enhance-and-segment pipeline, which combined Hessian-based filtering and graph cut segmentation for proximal femur segmentation in CT.
Gao~\textit{et al}.~\cite{gao_2022_Deep} introduced a multi-angle projection network for rib fracture segmentation in CT, utilizing rib extraction, fracture segmentation, and multi-angle projection fusion modules to capture rib and fracture features.
Liu~\textit{et al}.~\cite{liu_2023_Pelvic} developed a two-stage approach for pelvic fracture segmentation, using a pelvic bone segmentation network followed by a fracture segmentation network.
Cai~\textit{et al}.~\cite{cai_2024_Automatic} proposed a 3D U-Net-based method for knee CT segmentation, generating 3D fracture maps and aiding Schatzker classification to improve clinical diagnostic efficiency.
Zhou~\textit{et al}.~\cite{zhou_2024_CrossScale} designed a framework for hip fracture segmentation, incorporating a cross-scale attention mechanism and a surface supervision strategy to enhance fracture representation and boundary accuracy.

Some studies have focused on semi-supervised segmentation of bone structures.
Zhao~\textit{et al}.~\cite{zhao_2019_SemiSupervised} proposed a self-training method for finger bone segmentation in hand X-rays, utilizing a U-Net model with a conditional random field module to generate pseudo-labels for unlabeled images.
Burton~\textit{et al}.~\cite{burton_2020_Semisupervised} introduced a teacher-student model for knee MRI segmentation, employing a Monte Carlo patch sampling strategy to improve accuracy.
Liu~\textit{et al}.~\cite{liu_2020_Semisupervised} explored generative adversarial network (GAN)-based methods for lumbosacral structure segmentation on thin-layer CT, using semi-cGAN and few-shot-GAN models.
Li~\textit{et al}.~\cite{li_2023_Patchshufflebased} proposed a patch-shuffle-based augmentation technique, leveraging bone structure uniqueness and consistency loss to align features between original and shuffled CT slices.
Li~\textit{et al}.~\cite{li_2025_Semisupervised} developed a semi-supervised tooth segmentation method combining an entropy-guided mean-teacher approach and a weakly mutual consistency network.

\section{Methods}
\subsection{Network Architecture}

\begin{figure}[t]
	\begin{center}
		\includegraphics[width=\linewidth]{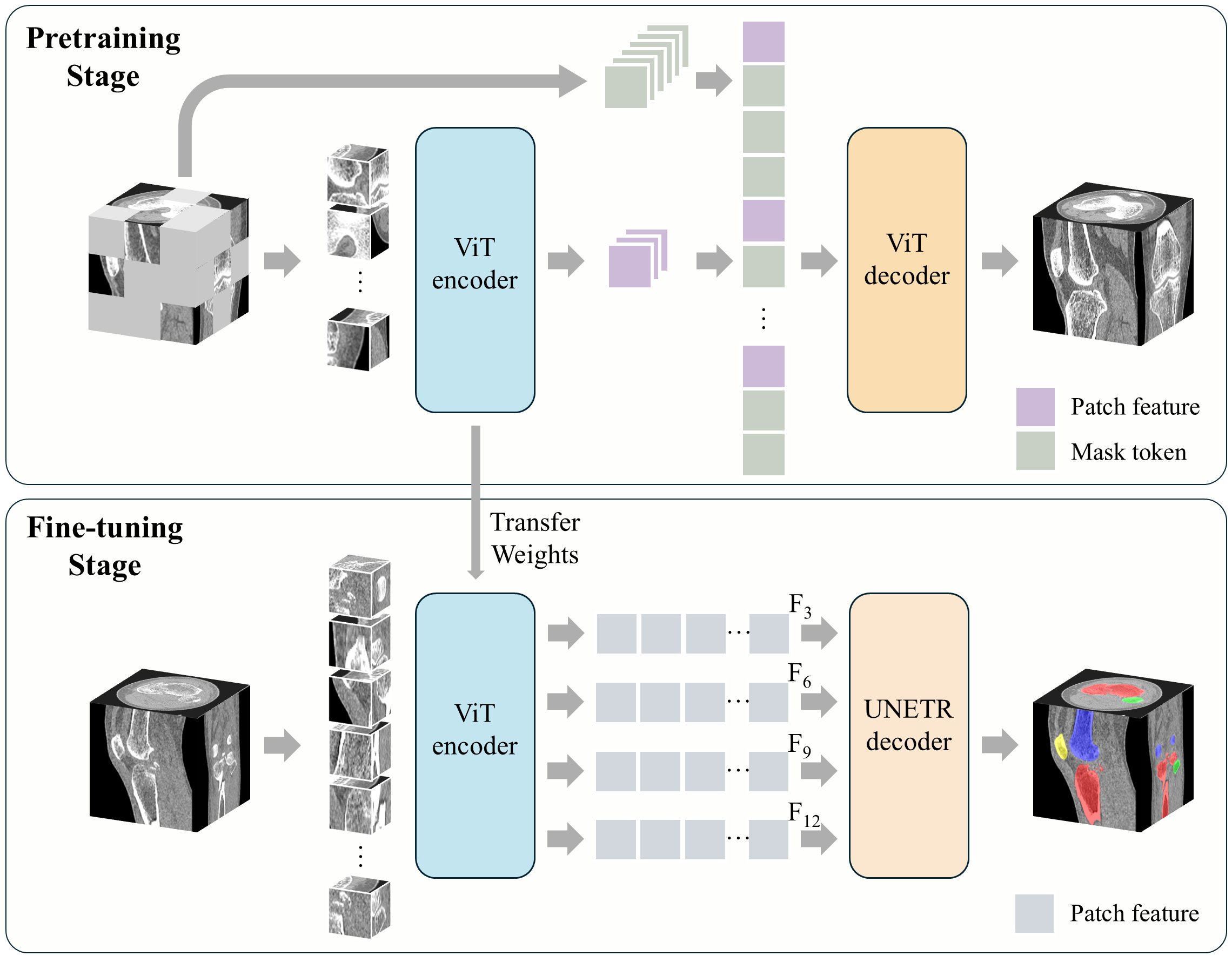}
	\end{center}
	\caption{The proposed pretrain and finetune framework for TPF segmentation in CT. Note that the pretrain stage uses unlabeled data while the finetune stage leverages limited labeled data.}
	\label{fig:network}
\end{figure}

As illustrated in Fig.~\ref{fig:network}, the proposed segmentation network comprises two stages: a MAE~\cite{he_2022_Masked} pretraining stage and an UNETR~\cite{hatamizadeh_2022_UNETR} fine-tuning stage.
The MAE employs an asymmetric encoder-decoder design.
Its Vision Transformer (ViT)~\cite{dosovitskiy_2021_Image} encoder processes only visible tokens, while a lightweight decoder reconstructs masked patches using the patch-wise output from the encoder and trainable mask tokens.
The UNETR follows the U-Net~\cite{ronneberger_2015_UNet} principle of skip connections, linking features from multiple encoder resolutions to the decoder.
The UNETR decoder takes a sequence of representations from the encoder, reshapes them to restore spatial dimensions, and progressively upsamples and concatenates them with shallower features to achieve high-resolution segmentation outputs.
The detailed structures of the MAE and UNETR are described in the following subsections.

\subsection{Pretrain stage with MAE}
Masked image modeling (MIM)~\cite{he_2022_Masked,wei_2022_Masked,xie_2022_SimMIM,liu_2023_MixMAE} is widely used in self-supervised learning because of its simple design and strong performance.
The main idea of MIM is to mask some image patches at the input and reconstruct them at the output.
This helps the network learn to predict the missing parts by using information from the surrounding context.
We believe this ability to use contextual information is important for bone CT scans, as it helps capture the overall bone structures and the details of fractures.

\begin{figure}[t]
	\begin{center}
		\includegraphics[width=\linewidth]{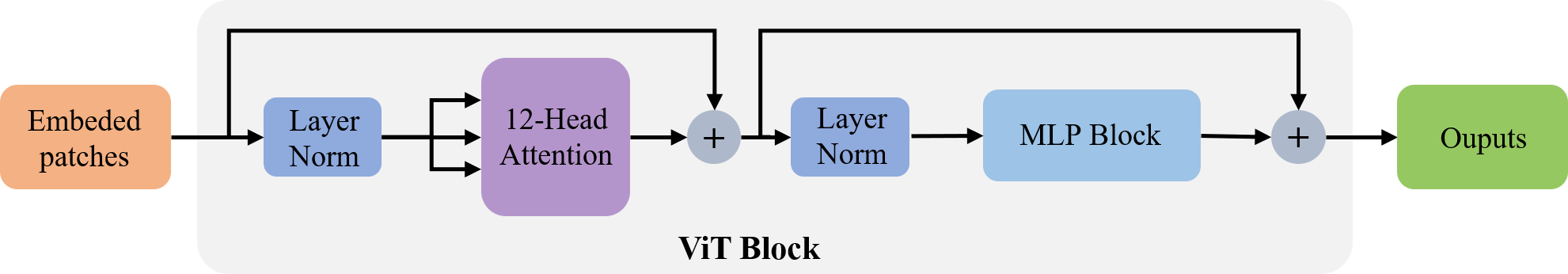}
	\end{center}
	\caption{Architecture of the ViT block used in the MAE.}
	\label{fig:vit}
\end{figure}

Among the different MIM methods, MAE is both simple and effective.
The encoder of MAE consists of 12 identical ViT blocks, each containing a multi-head self-attention layer and a fully connected feedforward network (see Fig.~\ref{fig:vit}).
The input image is divided into patches of size 16$\times$16$\times$16, which are linearly projected and flattened into one-dimensional vectors. Positional encoding is added to these patches to retain spatial information. 
Among these patches, 75\% are masked, leaving only the visible patches to be processed by the ViT blocks.
Similar to the encoder, the decoder in MAE comprises 8 layers of ViT blocks.
The decoder's input includes all patches, which consist of visible patches output from the encoder and masked patches replaced with mask tokens.
Each mask token is a shared, learnable vector representing the patch to be reconstructed.
Positional encoding is also added to the decoder's input.
The decoder's final layer is a linear projection whose output channels correspond to the number of pixel values in a patch.
The output of the decoder is reshaped to form a reconstructed image. 
The training loss function computes the mean squared error (MSE) between the reconstructed and original images in the pixel space.
As with most MIM strategies, the reconstruction loss is computed only on the masked patches.

Through this self-supervised reconstruction task, the encoder learns both the global structures and detailed features of the bone from the contextual information, which effectively benefits subsequent segmentation tasks.

\subsection{Finetune stage with UNETR}
U-Net has been widely used in medical image segmentation tasks, achieving favorable results across various applications due to its ``U-shaped'' architecture for the encoder and decoder.
Similarly, UNETR adopts this successful ``U-shaped'' design, utilizing ViT blocks as the encoder, which makes it naturally compatible with ViT encoders pretrained using the MAE strategy.
In UNETR, the decoder receives a sequence of feature representations from the encoder.
Specifically, the outputs from the 3rd, 6th, 9th, and 12th layers of the encoder during the pretraining phase, along with the original image, are utilized as inputs to the decoder side of UNETR. 
These feature representations are reshaped to restore their spatial dimensions and then progressively upsampled.
They are further concatenated with features from shallower layers to generate higher-resolution segmentation outputs.

\begin{figure}[t]
	\begin{center}
		\includegraphics[width=\linewidth]{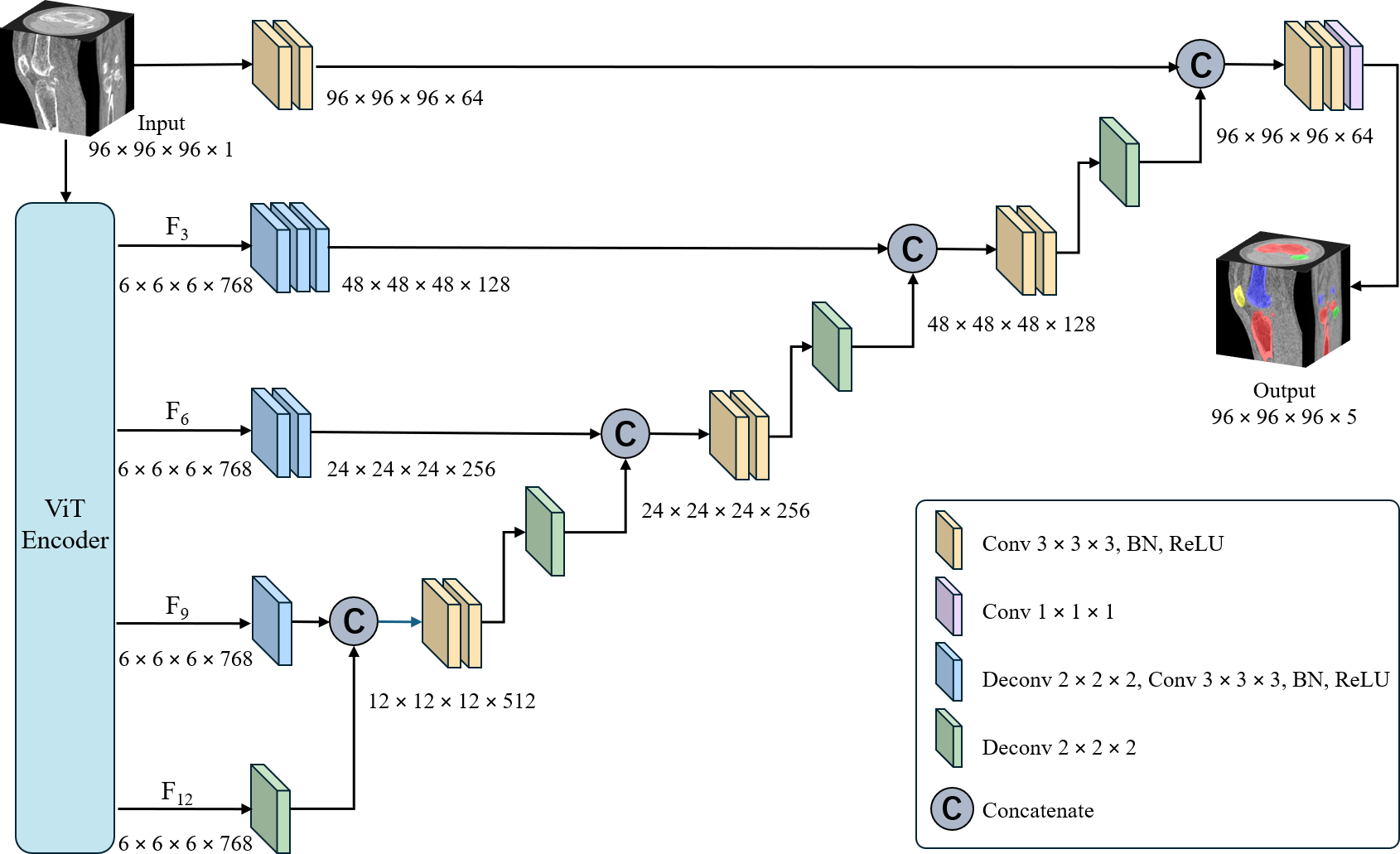}
	\end{center}
	\caption{Architecture of the UNETR.}
	\label{fig:unetr}
\end{figure}

The detailed architecture of the UNETR employed in our method is illustrated in Fig.~\ref{fig:unetr}.
The segmentation task is optimized using a combined loss function comprising Dice loss and cross-entropy loss for supervision.

\section{Experiments and Results}
\subsection{Materials}
Experiments were mainly carried on CT scans obtained from 180 patients with TPF.
All these data were collected from the Shenzhen Second People's Hospital.
This study was conducted retrospectively and therefore the research received a waiver of approval from our institutional review board.
In addition, we further employed a public pelvic CT dataset~\cite{work8} containing 103 scans with hip fractures (HF) to test the transferability of our method.

For preprocessing, voxels with CT values outside the range of -500 to 500 HU were excluded using a coarse filter, followed by windowing and centering to enhance and normalize the CT volumes.
The TPF dataset was resampled to a voxel spacing of 0.328$\times$0.328$\times$1.000mm$^3$, and the HF dataset was resampled to 0.839$\times$0.839$\times$0.800mm$^3$.
The TPF dataset was split into 20 labeled and 50 unlabeled training cases, 10 validation cases, and 100 testing cases.
Similarly, the HF dataset was divided into 20 labeled training cases, 10 validation cases, and 73 testing cases.

\subsection{Comparison and Evaluation}
We compared our method with mainstream semi-supervised approaches under varying labeled data proportions.
The baseline followed the student-teacher co-training paradigm, where the teacher generated pseudo-labels for weakly augmented unlabeled data, and the student was trained on weakly augmented labeled data (with ground truth) and strongly augmented unlabeled data (with pseudo-labels).
For the TPF dataset, strong augmentations comprised random masking, rotations, scaling, and translations.

\begin{table}[t]
	\centering
	\caption{The numerical segmentation results of different methods for tibial plateau fractures (Mean$\pm$SD, best results are highlighted in bold). }
	\label{tab:results1}
	\resizebox{\linewidth}{!}{
		\begin{tabular}{c|c|c|c|c}
			\hline
			Methods & Labeled data & DSC (\%) $\uparrow$ & ASSD (mm) $\downarrow$ & 95HD (mm) $\downarrow$ \\
			\hline
			\multirow{3}{*}{\makecell{Semi- \\ supervised}} 
			& 20 & 93.54$\pm$6.71 & 3.15$\pm$4.36 & 17.31$\pm$22.57 \\
			& 10 & 90.68$\pm$7.44 & 3.19$\pm$3.05 & 18.47$\pm$17.92 \\
			& 5  & 87.61$\pm$9.60 & 7.67$\pm$5.16 & 51.02$\pm$26.94 \\
			\hline
			\multirow{3}{*}{Ours} & 20 & \textbf{95.81$\pm$5.93}  & \textbf{1.91$\pm$3.64}   & \textbf{9.42$\pm$18.54} \\
			& 10 & 94.15$\pm$6.57   & 3.51$\pm$5.91   & 14.83$\pm$20.11 \\
			& 5  & 92.90$\pm$7.88   & 6.62$\pm$6.52   & 32.31$\pm$27.52 \\
			\hline
		\end{tabular}%
	}
\end{table}

\begin{table}[t]
	\centering
	\caption{The numerical segmentation results of different methods for hip fractures (Mean$\pm$SD, best results are highlighted in bold). }
	\label{tab:results2}
	\resizebox{\linewidth}{!}{
		\begin{tabular}{c|c|c|c|c}
			\hline
			Methods & Labeled data & DSC (\%) $\uparrow$ & ASSD (mm) $\downarrow$ & 95HD (mm) $\downarrow$ \\
			\hline
			\multirow{3}{*}{\makecell{U-Net}} 
			& 20 & 90.94$\pm$3.11 & 4.37$\pm$3.21 & 32.44$\pm$26.41 \\
			& 10 & 59.37$\pm$2.97 & 16.06$\pm$3.22 & 82.96$\pm$18.86 \\
			& 5  & 44.77$\pm$2.64 & 37.70$\pm$6.54 & 113.52$\pm$13.01 \\
			\hline
			\multirow{3}{*}{\makecell{UNETR}} 
			& 20 & 88.71$\pm$4.36 & 15.53$\pm$5.51 & 94.13$\pm$24.79 \\
			& 10 & 70.09$\pm$6.89 & 38.25$\pm$5.60 & 138.87$\pm$12.39 \\
			& 5  & 52.46$\pm$5.73 & 42.96$\pm$5.10 & 148.95$\pm$14.00 \\
			\hline
			\multirow{3}{*}{Ours} & 20 & \textbf{96.75$\pm$2.12} & \textbf{1.99$\pm$3.02} & \textbf{10.84$\pm$23.76} \\
			& 10 & 93.16$\pm$3.60 & 5.33$\pm$3.52 & 43.80$\pm$30.61 \\
			& 5  & 89.87$\pm$3.63 & 14.17$\pm$4.73 & 72.23$\pm$29.07 \\
			\hline
		\end{tabular}%
	}
\end{table}

To further investigate the transferability of our method across different datasets, we pretrained the encoder using MAE on the TPF dataset and finetuned it for HF segmentation task.
We compared the performance of our approach with models trained exclusively on HF data, including UNETR and U-Net.

The evaluation metrics included the Dice similarity coefficient (DSC) for volumetric overlap, average symmetric surface distance (ASSD) for surface distance, and symmetric 95\% Hausdorff distance (95HD) for localized disagreement.
DSC measures the overlap between segmented regions and ground-truth masks, while ASSD evaluates the distance between segmented and ground-truth surfaces.
95HD, more sensitive to localized errors, represents the 95th percentile of Hausdorff distances.
Better segmentation results are indicated by higher DSC and lower ASSD and 95HD values.

\subsection{Experimental Setup}
We implemented our method in PyTorch\footnote{\url{https://pytorch.org/}} and MONAI\footnote{\url{https://monai.io/}}. 
All models were trained using NVIDIA RTX 2080Ti GPUs.
We used the Adam optimizer with a batch size of 2 for both pretraining and fine-tuning, along with a cosine annealing learning rate scheduler. 
During pretraining, the initial learning rate was set to 6.4e-3, and training was conducted for up to 10,000 epochs. 
For fine-tuning, the initial learning rate was adjusted to 3.44e-2, and training was limited to 5,000 epochs. 
Additionally, a layer-wise learning rate decay strategy (decay factor=0.75) was applied during fine-tuning to stabilize the pretrained encoder, with deeper layers assigned smaller decay rates.
Semi-supervised baseline method was first trained in a fully supervised manner for 5,000 epochs, followed by an additional 10,000 epochs of semi-supervised training.

\subsection{Results}

\begin{figure}[t]
	\begin{center}
		\includegraphics[width=\linewidth]{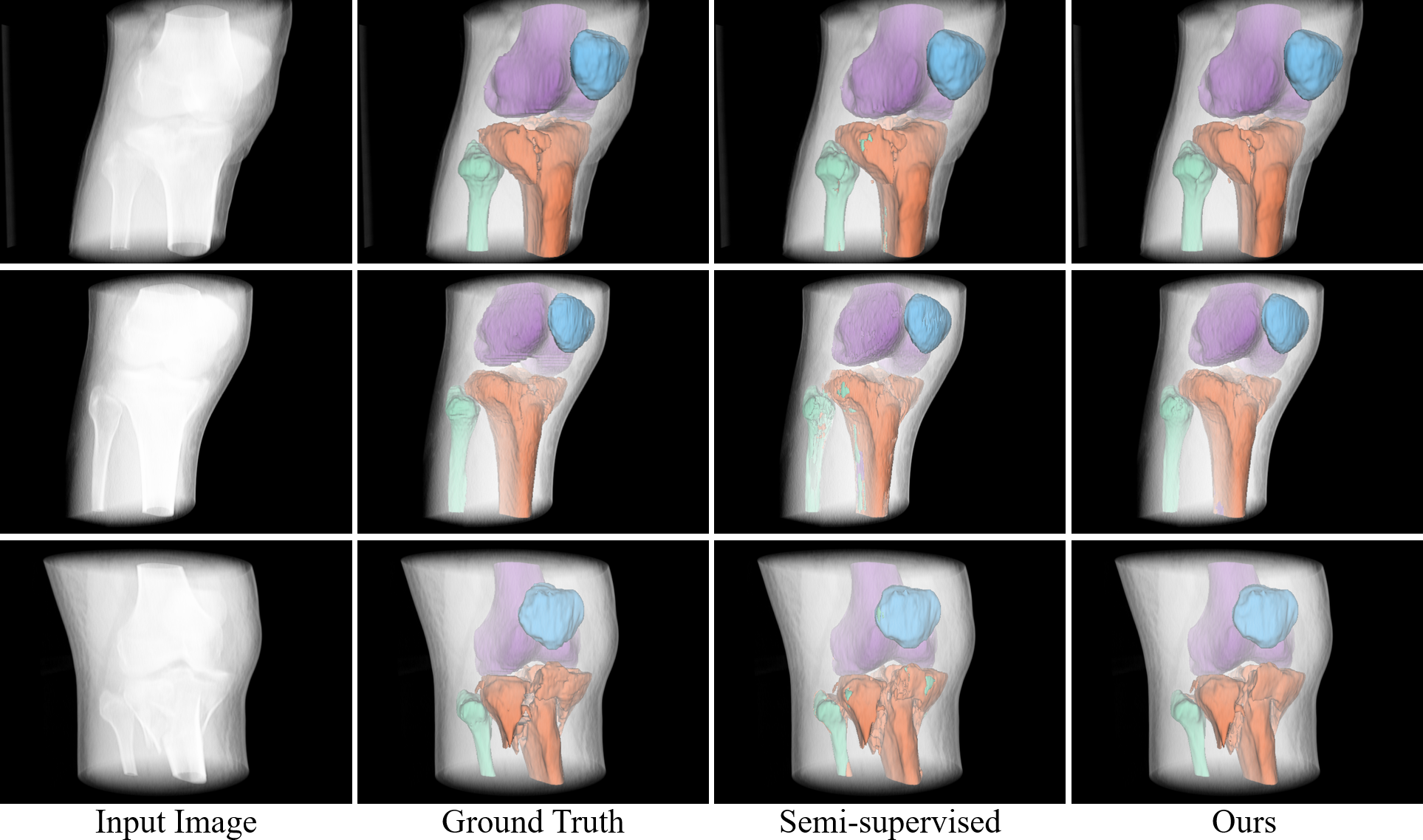}
	\end{center}
	\caption{3D visualization of the segmentation results for tibial plateau fractures.}
	\label{fig:3d_tibia}
\end{figure}

\begin{figure}[t]
	\begin{center}
		\includegraphics[width=\linewidth]{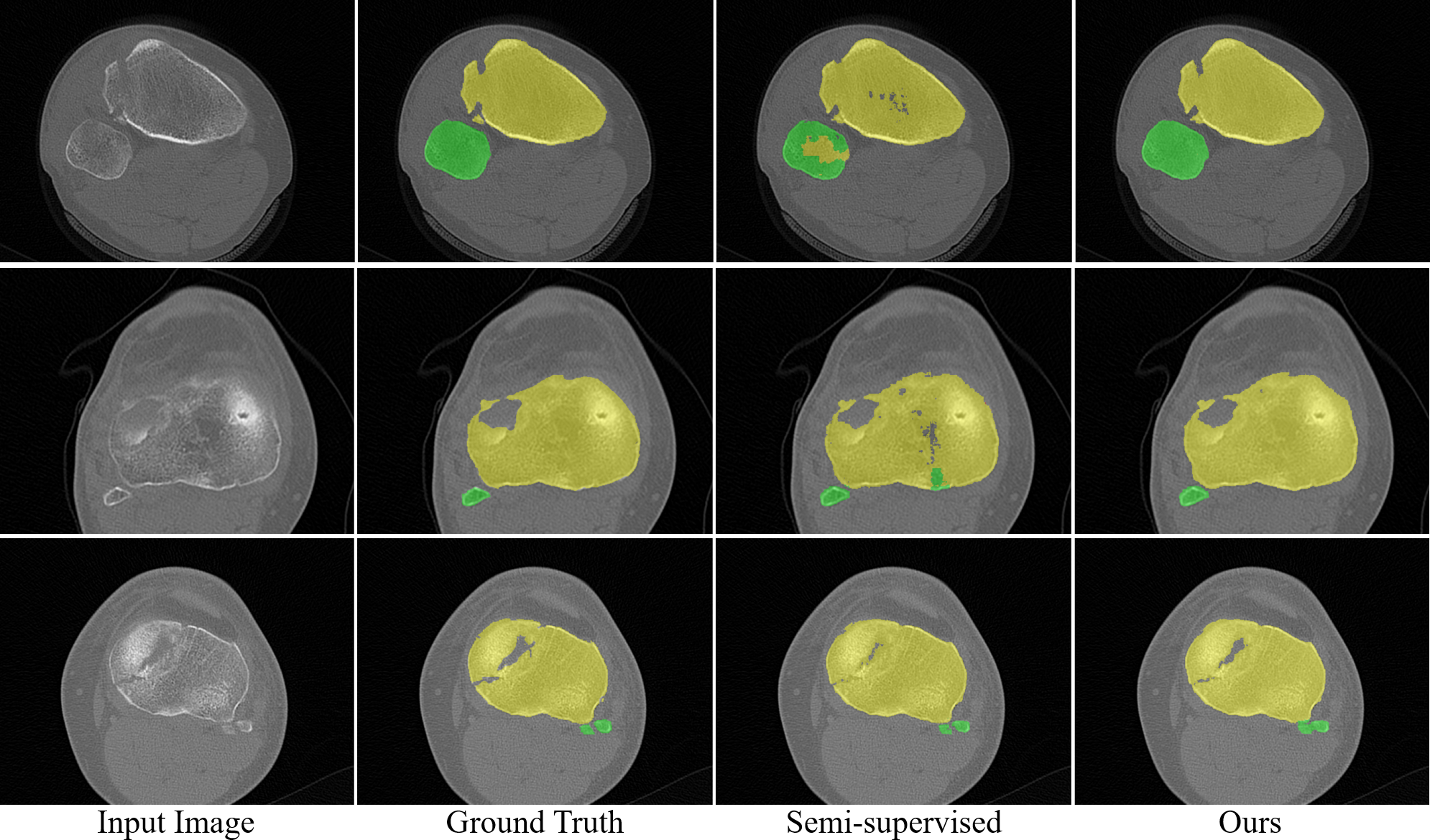}
	\end{center}
	\caption{2D visualization of the segmentation results for tibial plateau fractures.}
	\label{fig:2d_tibia}
\end{figure}

The TPF segmentation results are summarized in Table~\ref{tab:results1}.
As demonstrated, our method consistently outperforms other semi-supervised approaches across all evaluation metrics, highlighting its superior performance.
With only five labeled cases, our method achieves an average DSC of 92.90\%, an ASSD of 6.62mm, and a 95HD of 32.31mm, showcasing its effectiveness even with minimal labeled data.
Notably, when trained with just 10 labeled cases, our method achieves an average DSC of 94.15\%.
Compared to training with 20 labeled cases, the DSC decreases by only 1.36\%, further emphasizing that our approach not only delivers high segmentation accuracy with limited annotations but also demonstrates strong potential for practical applications in low-data scenarios.

Fig.~\ref{fig:3d_tibia} shows the 3D visualization of segmentation results between semi-supervised learning and our method.
It is obvious that our method provides more accurate segmentation results.
Fig.~\ref{fig:2d_tibia} further visualizes some detailed 2D comparisons, showing that our method produces segmented boundaries most similar to the ground truth.
Even in cases with complex fracture morphologies, our method demonstrates a strong ability to accurately segment skeletal structures.

Table~\ref{tab:results2} compares the segmentation performance of different methods on the HF dataset.
Our method, pretrained on unlabeled TPF data using MAE and fine-tuned on the HF data, achieves the best segmentation results under the same amount of labeled data.
This further highlights that, compared to semi-supervised learning, the MAE reconstruction task enables the model to learn more general and transferable skeletal features, leading to improved performance in cross-dataset scenarios.

\section{Conclusion}
This study presents an accurate and automated method for the segmentation of tibial plateau fractures in CT scans.
The primary motivation is to leverage the distinct features of bone structures, particularly the anatomical and fracture characteristics, to enhance segmentation performance.
We propose a pretrain-finetune strategy based on MAE, where the pretraining stage captures global anatomical structures and fine-grained fracture details from unlabeled data, and the fine-tuning stage uses limited labeled data to guide segmentation. 
This approach effectively reduces dependence on extensive annotations while learning generalizable and transferable features.
Experiments on both in-house and public datasets prove the efficacy of the proposed method.
Compared with existing semi-supervised approaches, our method demonstrates superior performance and strong transferability across different datasets, highlighting its potential for broader applications in fracture segmentation and other medical imaging tasks.

\FloatBarrier
\bibliographystyle{IEEEtran}
\bibliography{scholar}
\end{document}